\documentclass[12pt]{article}

\title{Community Detection with Metadata in a Network of Biographies of Western Art Painters}

\usepackage[a4paper,margin=2cm,marginparwidth=1cm]{geometry}
\usepackage{hyperref}

\usepackage[round]{natbib}

\usepackage{amsmath}
\usepackage{graphicx}

\usepackage{caption}
\usepackage{subcaption}
\usepackage{float}

\usepackage{geometry}

\usepackage{amsmath,amssymb,amscd}

\usepackage{hyperref}

\providecommand{\href}[2]{\texttt{#2}}
\providecommand{\url}[1]{\texttt{#1}}

\newcommand{\Ccal}{\mathcal{C}}
\newcommand{\Ncal}{\mathcal{N}}

\usepackage{graphicx}

\usepackage{amsmath,amssymb,amscd}

\newcommand{\secref}[1]{Section~\ref{#1}}

\usepackage{setspace}

\begin{document}

 \vspace*{0.5cm}

\begin{center}
{\Large\textbf{Community Detection with Metadata in a Network of Biographies of Western Art Painters}} \\[0.5cm]
 {\large {Michael Kitromilidis},
  \href{http://www.imperial.ac.uk/people/t.evans}{Tim S.\ Evans}}
 \\[0.5cm]
\href{http://complexity.org.uk/}{Centre for Complexity Science}, and \href{http://www3.imperial.ac.uk/theoreticalphysics}{Theoretical Physics Group},
Imperial College London, SW7 2AZ, U.K.
 \\[0.5cm]  Key Words: Social Influence, Art History, Wikipedia Networks, Community Detection
\end{center}

\begin{abstract}

In this work we look at the structure of the influences between Western art painters as revealed by their biographies on Wikipedia. We use a modified version of modularity maximisation with metadata to detect a partition of artists into communities based on their artistic genre and school in which they belong. We then use this community structure to discuss how influential artists reached beyond their own communities and had a lasting impact on others, by proposing modifications on standard centrality measures.
\end{abstract}

%

\section{Introduction}

Many studies in network science are concerned with community detection, proposing various methods and algorithms both for the classification of nodes into clusters, and for the evaluation of such classifications \citep{Dan05,Fort10,Fort16,New06}. Particularly in social networks, the evaluation of a community detection method is concerned with the nature of the clusters into which nodes are placed, for example the social groups, professions or even special interests that the nodes in each cluster share  \citep{Ahn10,Bar02,Eva09,Eva10,Gle03,Gold15}. Recent advances in the literature suggest evaluating a clustering in terms of a ground truth, which is based on metadata, characteristics that nodes possess which are external to the structure and topology of the network \citep{Hric14,Hric16,Peel17,Yang13}.

In this work we aim to build on the existing framework for community detection with metadata and propose a new area where this methodology can be applied, by forming a network of biographical connections between Western art painters in a timespan ranging from the 14$^{th}$ to the 20$^{th}$ centuries. We perform community detection with the aim of matching identified clusters to artistic movements. Furthermore, we use the community structure(s) of the network to re-define standard centrality and brokerage measures in order to highlight painters, whose links to artistic movements beyond their own, can classify them as being influential.

This paper is organised as follows.  In \secref{sdata} we introduce the empirical dataset we will be using in later analysis. In \secref{scommunity} we discuss community detection and introduce our two alternative measures for assessing a community partition given metadata information. We then test these measures in three cases: the classic example of Zachary's Karate Club, a synthetic network we are producing, and our empirical network of painters. In \secref{scentrality} we introduce variations on standard centrality measures taking into account an underlying community structure. We see how the standard partitions and the partitions motivated by our measures help us highlight nodes which have a bridging role across communities and present examples from our painter network.

\section{Network Definition and Properties}\label{sdata}

\subsection{Data sources and network definition}

The context of our network comes from Art History, as we build a network of Western art painters. In this section we introduce the network and some of its main properties.

We collect data from the \href{http://www.wga.hu}{Web Gallery of Art}, an online repository of more than 40,000 artworks, which is freely accessible for education and research, for two main purposes: firstly to specify a concrete list of which painters to include in our network and secondly to collect metadata for those painters. Wikipedia also contains metadata for some artists (Wikidata), but we only select that information from the WGA because it is a complete set of metadata for all artists in our database.

From the list of artists, we are finding the corresponding Wikipedia page for each artist, using the Wikipedia Python API. As we are focusing on painter collaboration, we create a network for the painters present in the database, who we want to link according to the encounters they may have had with each other. The dataset can be found online \citep{Kit17}.

The nodes in our network are individual painters and edges between nodes are drawn according to biographical connections between artists, which may correspond to influence or other social and contextual links. To quantify that, we draw an edge when the Wikipedia page of one artist links to another artist in the database. In doing so we construct a network of $N = 2474$ nodes and $E = 9568$ edges. This is a simple, unweighted and undirected network.

The reason why we are choosing an undirected network is because of the way we are drawing the edges; a Wikipedia page of an artist may refer both to the artists that influenced a painter but also the ones that were influenced by them, the ones with whom they shared a workshop, or even painters whose works they may have collected and owned. Weight and multiplicity in edges may be more appropriate, as certainly some artists may be more closely connected than others or might have a greater influence; however this is something that is not straightforward to determine with this data extraction process.

Some further manual cleaning-up of the data is required, as some nodes are duplicate pages of artists, or may correspond to other kinds of artists (e.g. sculptors) but not painters, due to the structure of our data sources. After the cleaning-up and the further removal smaller isolated components (typically singletons or two nodes only connected with each other which were not relevant for our analysis), we are left with a graph of $N = 2113$ nodes and $E = 9417$ edges.

\subsection{Basic network description}

We begin our analysis with a short statistical description of the painter network. It has average degree $\langle k \rangle = 8.9$, clustering coefficient $\langle C \rangle = 0.28$, average shortest path length $\langle l \rangle = 4.07$ and diameter $d=14$. Figure \ref{img1} is a plot of the complementary cumulative degree distribution, which exhibits a truncated power-law behaviour (more in the next section). Degree-degree correlations present no significant features beyond a weak positive correlation $\rho = 0.26$.

It is interesting to note that the most highly connected artists in the network are largely well known names, see Table \ref{deg}. The best connected artist turns out to be \href{https://en.wikipedia.org/wiki/Peter_Paul_Rubens}{Rubens}, a master of the Baroque age. We also see many masters of Renaissance art featuring on this list such as \href{https://en.wikipedia.org/wiki/Raphael}{Raphael}, \href{https://en.wikipedia.org/wiki/Titian}{Titian} and \href{https://en.wikipedia.org/wiki/Leonardo_da_Vinci}{Leonardo da Vinci}.

The fact that the best linked artists are very well known and influential, is to be expected. Many painters' Wikipedia pages will have references to some of the masters they were inspired by or whose workshops they might have worked in, and such connections generate many links to these more recognisable names. The only surprising occurrence in this list is the lesser known painter \href{https://en.wikipedia.org/wiki/Karel_van_Mander}{Karel van Mander}, who does however have many connections to other painters due to the fact that he was mainly an art historian and their biographer. This observation is perhaps an good illustration of a characteristic feature that one might expect from our data collecting process.

\begin{table}
\centering
\caption{\textbf{Most connected painters in the network ranked by degree.}}
\begin{tabular}{cc}
\hline
\textbf{Painter} & \textbf{Degree} \\
\hline
Peter Paul Rubens & 154 \\
Rembrandt & 146 \\
Raphael & 123 \\
Caravaggio & 118 \\
Titian & 101 \\
Giorgio Vasari & 96 \\
Diego Velazquez & 94 \\
Karel van Mander & 85 \\
Michelangelo & 83 \\
Leonardo Da Vinci & 74 \\
\hline
\end{tabular}
\label{deg}
\end{table}

\subsubsection{The degree distribution}

\begin{figure}[h]
\centering
\includegraphics[width=0.5\textwidth]{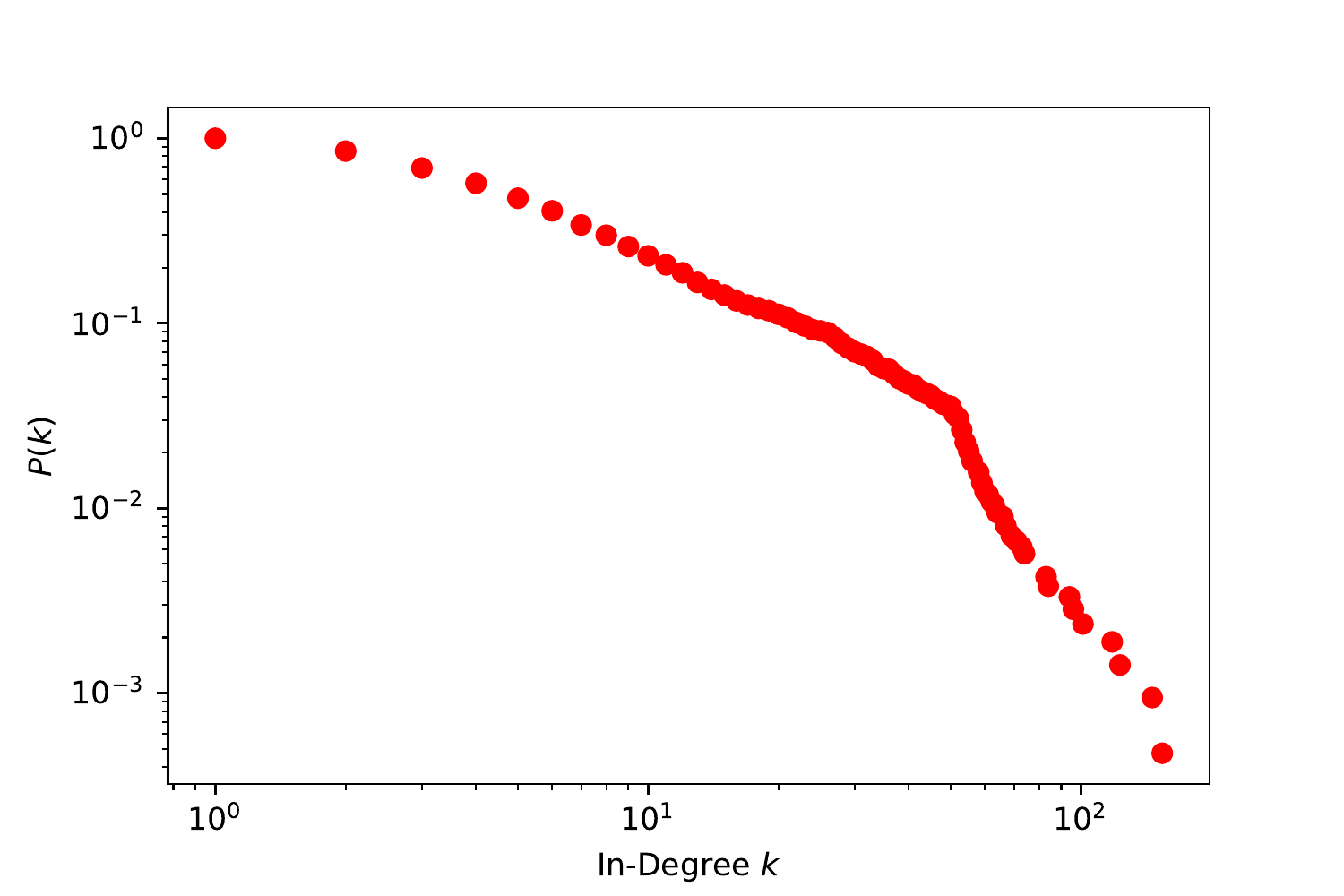}
\caption{Complementary cumulative degree distribution for the painter network.}
\label{img1}
\end{figure}

The degree distribution of the painter network (Figure \ref{img1}) displays an interesting, and slightly uncommon ``knee"-like distribution. A distribution like this, which is close to a truncated power law with two exponents (data fitting shows a good fit of $\gamma\approx 2.4$ to $k\leq 60$) has been observed in a few other artificial systems - most notably it has occurred in some transportation systems \citep{Ama00,Gui05,Hu09}, in those contexts corresponding to phenomena related to capacity constraints.

In the context of our painters network this behaviour can be explained partly due to the way that the Wikipedia articles in our database are written (showing a similar behaviour with the degree distribution in both the article length and number outgoing links distribution).

\subsubsection{Centrality measures}

Apart from the degree, one can also look at other centrality measures to see how the painters' connectivity ranks in the network. Table \ref{betcent} shows the ranks of some of the most popular centrality measures.

\begin{table}
\centering
\caption{\textbf{Centrality measures in the painter hyperlink network.}}
\begin{tabular}{ p{10mm}  p{30mm}  p{30mm}  p{30mm}  p{30mm} }
\hline
\textbf{Rank}  & \textbf{Betweenness} & \textbf{Closeness}  & \textbf{Eigenvector} & \textbf{Pagerank} \\
\hline
1& Rembrandt & Rubens  & Caravaggio & Rembrandt \\
2 & Rubens & Rembrandt & Simon Vouet & Raphael  \\
3 & Raphael  & Raphael  & Artemisia Gentileschi & Rubens \\
4 & Titian & Titian & Jusepe de Ribera  & Vasari  \\
5 & Caravaggio & Caravaggio  & Cecco del Caravaggio  & Titian \\
\hline
\end{tabular}
\label{betcent}
\end{table}

\begin{figure}[h]
\centering
\includegraphics[width=0.7\textwidth]{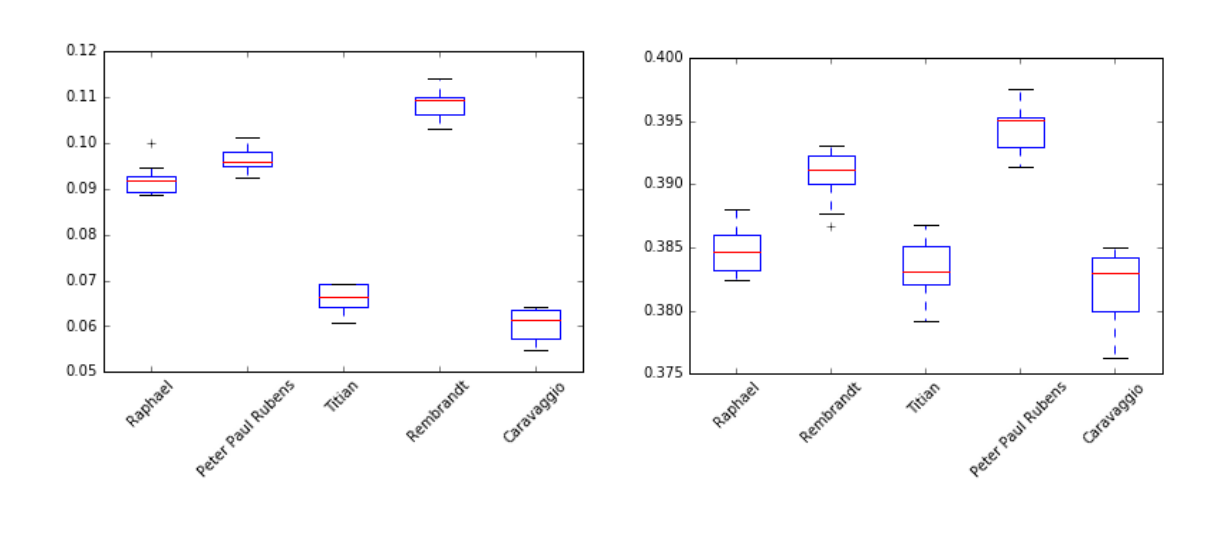}
\caption{Betweenness centrality (left) and closeness centrality (right) for the top ranked painters in each measure, in randomly perturbed networks with 5\% rewiring. The plots are typical box-and-whisker plots, with the width of the box corresponding to the first and third quartiles of the randomised network centrality values.}
\label{centralities1}
\end{figure}

Figure \ref{centralities1} shows some tests of the robustness of these centrality measures, after some randomised perturbation of the network (5\% edge rewiring) as done in the work by \cite{Che17}.

The centrality measures in this case reveal similar results as the degree, effectively highlighting the same, highly recognisable painters with some minor reordering. The most interesting result comes from the eigenvector centrality measure, which highlights the painter \href{https://en.wikipedia.org/wiki/Caravaggio}{Caravaggio} and many artists close to him (in fact he developed his own sub-movement, Caravaggism).

In order to understand more subtle kinds of artistic influence, we propose generalising centrality measures taking community structure into account, in \secref{scentrality}.

\section{Community detection with node attributes}\label{scommunity}

\subsection{Definition of Communities}

In this paper we will form communities of painters such that every painter is in one and only one community.  However we will also ask that communities are more tightly knit than average, so a typical community has more edges within the community than edges linking it to nodes outside the community. 

The first part of our definition is what is formally known as a partition of the set of nodes, the set of painters in our case. Let $\Ncal$ be the set of nodes (the painters) in our network.  Then each individual community is a non-empty subset $\Ccal_\alpha$ of painters, $\Ccal_\alpha \subseteq \Ncal$, such that there is no painter in two communities and all painters are in one unique community, formally $\Ccal_ \alpha\neq\emptyset$, $\Ccal_ \alpha \bigcap \Ccal_ \beta = \emptyset$ and $\Ncal =\bigcup_\alpha \Ccal_\alpha$.

The second aspect, that communities represent tight knit clusters within the network, is less precise and it is not surprising that there are many formal definitions and methods available to define this aspect. Here we will focus on one widely used method, the Louvain method \cite{Blo08}, whose results are shown in the Supplementary Information and visualised in Figure \ref{communities}. This method seeks to assign each node to a community in such a way that it gives an approximate maximum value for the modularity function $Q(A)$,

\begin{equation}
Q(A) = \frac{1}{2m} \sum_C \sum_{i,j\in C} \left(A_{ij}-\frac{k_ik_j}{2m}\right)
\label{mod}
\end{equation}

\begin{figure}[h]
\centering
\includegraphics[width=0.95\textwidth]{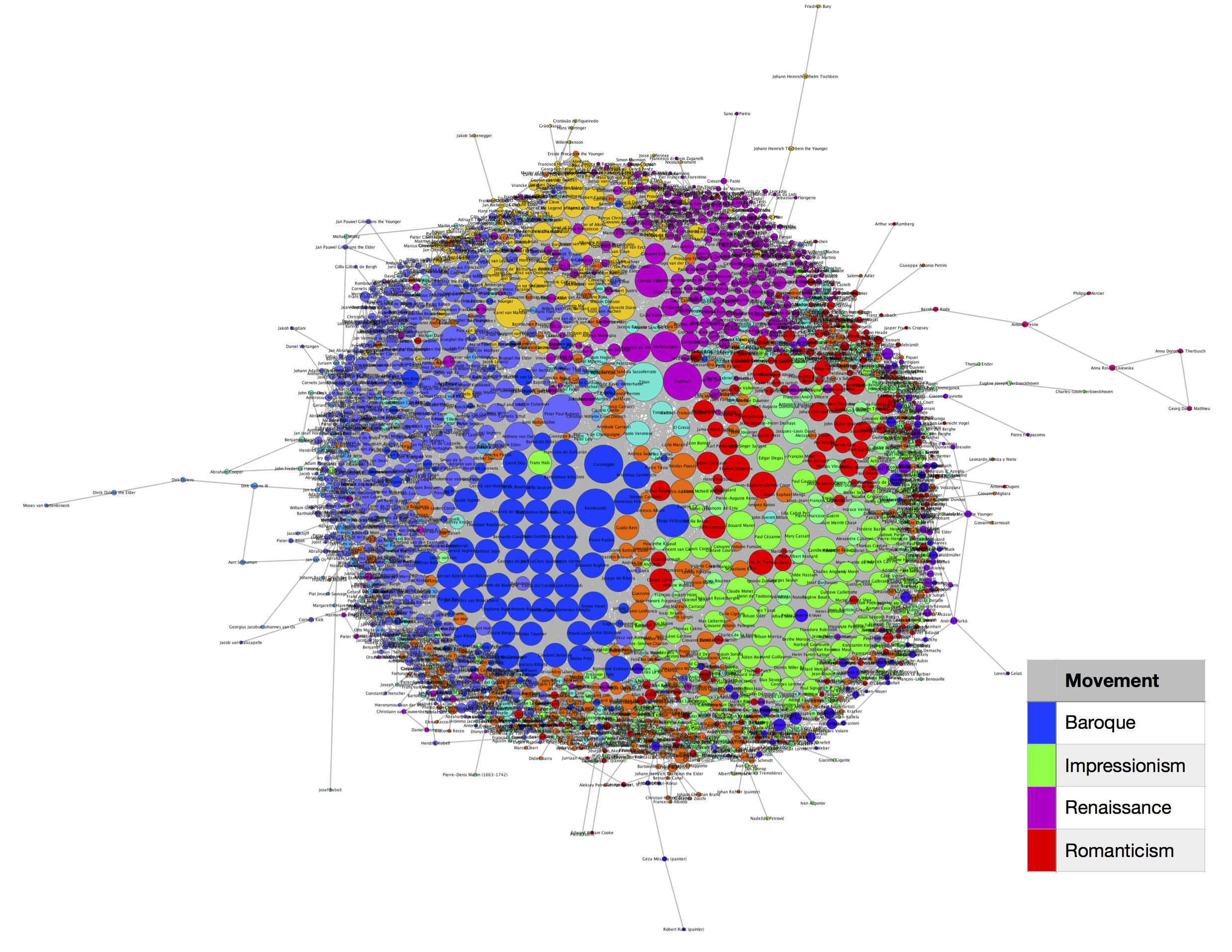}
\caption{Communities in the painter network; node size corresponds to the degree and colour to the community in which it is placed under the standard implementation of the Louvain method.}
\label{communities}
\end{figure}

We choose the Louvain method because it has been widely and successfully used while there are many fast numerical implementations available. In order to understand the nature of the clusters we look at the Wikipedia links that appear within each cluster and at the attributes gathered from the metadata; we expect, similarly to the work in \cite{Gle03} that the clusters will correspond to artistic movements. Some clusters do indeed show a clear correspondence to an artistic movement but not all of them do.

In practice what we find from our data is that our communities correspond to a mix of artistic movements and the location where the artists where primarily active. The results can be found in the Supplementary Information but we primarily observe that the algorithm does reveal considerably sized clusters for some of the most recognisable movements in Western art, the Renaissance, Baroque and Impressionism.

However, we also observe that some of the other identified communities are predominantly centred around a location, for example France or Italy. Motivated by this observation, we wish to force the Louvain method to look for communities at a finer scale. In the next section we formalise our approach.

\subsection{Nodes equipped with a set of attributes}

Using metadata together with a partition into communities is an idea which has been used in both earlier studies \citep{Dan05,New06,Ben01} and in more recent ones \citep{Peel17}. A common theme among the more recent advances in studies of communities with metadata is that metadata should not be used solely as an external ``ground-truth" which only assesses the quality of a partition, but instead information that can be used in conjunction with the network structure to detect more meaningful communities.

Traditional community detection methods only use the topological information in a network, the relations between nodes represented by edges. However, in the real world we usually have additional information - metadata. Often this is in the form of additional attributes, which can be either used post-hoc to test a method or, as we attempt to do here, used together with the network structure to detect communities.

To make this more precise and apply this rationale, we propose the following theoretical setup. We consider the situation where in a network of $N$ nodes, each node is equipped with a set of attributes $X_i = \left\lbrace x_1^{(i)},\ldots, x_m^{(i)}\right\rbrace$, where $x_a$ can take $k_a$ distinct values. This gives us a possibility of $\prod_{a=1}^m k_a$ combinations for different attribute configurations.

To assess the quality of a partition into communities $\mathcal{C} = \left\lbrace c_{\alpha},\ldots, c_{\nu}\right\rbrace$, we propose two alternative measures. An optimal partition in this context should isolate each specific configuration $\xi = \left(\xi_1,\ldots, \xi_m\right)$ in a single and unique cluster, and $\nu = \prod_{a=1}^m k_a$.

We therefore define the \textit{cluster homogeneity}
\begin{equation}
h_{\mathcal{C}}(c) = \frac{2}{|c|(|c|-1)}\sum_{i,j\in c} S(X_i,X_j)
\end{equation}
where the pre-factor is the number of possible pairs of nodes in the cluster $c_j$ and $S(x,y)$ is a suitable similarity measure. We also define the \textit{configuration entropy}
\begin{equation}
e_{\mathcal{C}}(\xi) = -\sum_{c_{\alpha}\in \mathcal{C}} p_{\alpha}(\xi)\log p_{\alpha}(\xi)
\end{equation}
where $p_{\alpha}(\xi)$ is the probability of finding the configuration $\xi$ in community $c_{\alpha}$, given by 
\begin{equation}
p_{\alpha}(\xi) = \frac{1}{|c_{\alpha}|}\sum_{i\in c_{\alpha}}\delta_{X_i,\xi}
\end{equation}

Informally, homogeneity is measured on a cluster of a partition and measures the similarity between the attributes of the nodes in that cluster. Entropy accepts as argument a specific configuration, and measures the fragmentation of this configuration into the various communities.

In an optimal partition (where each cluster has nodes of one attribute configuration and conversely each configuration belongs to a single cluster alone) homogeneity is equal to 1 and entropy is equal to 0. We further note the extreme values that these two measures can obtain: when the entire network is one community, $e = 0$ but $h\rightarrow 0$, whereas when every node is in its own community, $e=O(N)$ and $h = 1$.

\subsection{Implementation of quality measures}

We now test our theoretical measures in two examples as well as our empirical network and illustrate how they indicate whether a partition is too fine or two coarse.

\subsubsection{The Karate Club}

One of the most standard networks for testing methods in community detection is the Karate Club network studied by Zachary, and investigated by numerous approaches in network science. In this case each node in the network has one hidden attribute $X_i = 0$ or $1$, depending on whether the individual $i$ belongs in the faction of the manager or the instructor respectively.

The similarity function $S(x,y) = 1 - d_H(x,y)$, where $d_H(x,y)$ is the Hamming distance between $x$ and $y$; in this case simply 0 if the two nodes belong in the same faction and 1 if they belong in a different faction. The other parameters defined in Section 2.2 in this case are $m = 1$, $k_1 = 2$ and $\nu = 2$; i.e. an optimal partition should have two communities, one for each faction.

We see that the partition which maximises modularity (and detects four communities instead of the known two) fails to perform well on the measures of homogeneity and entropy. This is a case of \textit{overdetecting}, where and as expected the entropy value is quite high.

\begin{figure}[h]
\centering
\includegraphics[width=0.35\textwidth]{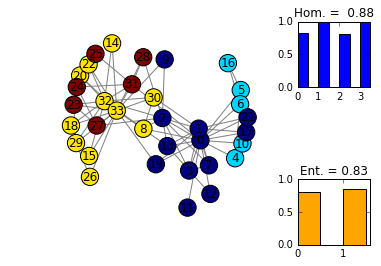}
\caption{Karate Club partition with modularity maximisation detects four communities (overdetection) and scores poorly on entropy.}
\end{figure}

\begin{figure}[h]
\centering
\includegraphics[width=0.35\textwidth]{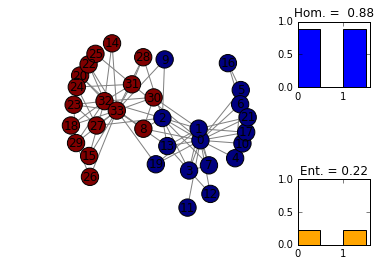}
\caption{Karate Club partition with merged clusters leaves the homogeneity unchanged but significantly reduces the entropy of the partition.}
\label{kc2}
\end{figure}

To overcome this issue we can merge identified communities, thus creating a coarser partition. After trying the several possible combinations of grouping the four clusters into fewer, we obtain the optimal way of splitting the network into only two communities (Figure \ref{kc2}). We note that this partition is also not perfect, but it is an improvement over the standard implementations of modularity maximisation (including those that can be obtained by tweaking the resolution parameter).

\subsubsection{Synthetic Network}

As a second, artificial example, we consider a network generated by the Stochastic Block Model \citep{Hol83}. Here we equip each node with a vector of two hidden attributes $X_i = (x_1^{(i)},x_2^{(i)})$, and each $x_j$ can take the value of $0$ or $1$ with equal probability; this means that the possible configurations are $\xi_1 = (0,0)$, $\xi_2 = (0,1)$, $\xi_3 = (1,0)$ and $\xi_4 = (1,1)$. More specifically here $m = 2$, $k_1,k_2 = 2$ and $\nu = 4$.

Two nodes are linked depending on the common attributes they share, i.e. they are linked with probability 1 if they have both attributes matching, with probability 1/2 if one attribute only is matching and are disconnected otherwise.

\begin{equation*}
P_{ij} = \left(
\begin{matrix}
1 & 1/2 & 1/2 & 0 \\
1/2 & 1 & 0 & 1/2 \\
1/2 & 0 & 1 & 1/2 \\
0 & 1/2 & 1/2 & 1
\end{matrix}
\right)
\end{equation*}

An optimal partition should uncover four communities in this case; however the standard implementation of modularity only yields two. In this case the average homogeneity is 0.75, as each community contains two kinds of nodes.

\begin{figure}
\centering
\includegraphics[width=0.35\textwidth]{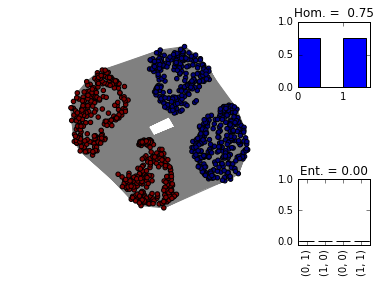}
\caption{Synthetic network partition with modularity maximisation; two communities detected (underdetection).}
\end{figure}

\begin{figure}
\centering
\includegraphics[width=0.35\textwidth]{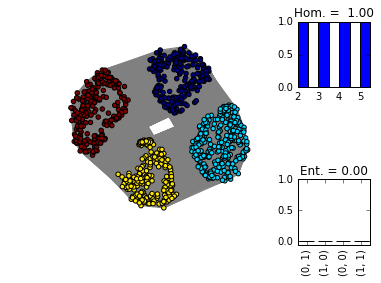}
\caption{Running modularity maximisation at the communities level uncovers the deeper level clusters and scores perfectly on homogeneity, while entropy remains at the optimal level.}
\label{sbm2}
\end{figure}

This is the opposite scenario from the Karate Club network, as we are \textit{underdetecting} communities; since $h<1$ and $e=0$ it means that our clusters contain more than one type of nodes. By running Louvain modularity maximisation again in each community (treated as a separate network) we are able to unfold the partition into four communities, as each original community splits into two (Figure \ref{sbm2}).

\subsubsection{The Painter Network}

We now implement our analysis on our empirical network of painters, looking for the partition at the fine level which optimises our two measures of homogeneity and entropy. Motivated the discussion in Section 2.1 we assign to each painter-node two attributes, their artistic movement and the country where they were working. These tags are also sourced from the WGA, and can be found in detail in the Supplementary Information.

In this case we have $m=2$ attributes, $k_1 = 11$ movements and $k_2 = 25$ locations. However as there is some overlap between some of the locations and most importantly some movement-location combinations are not realistic we should expect a considerably smaller number than $\nu = 275$ communities in an optimal partition.

\begin{figure}
\centering
\includegraphics[width=\textwidth]{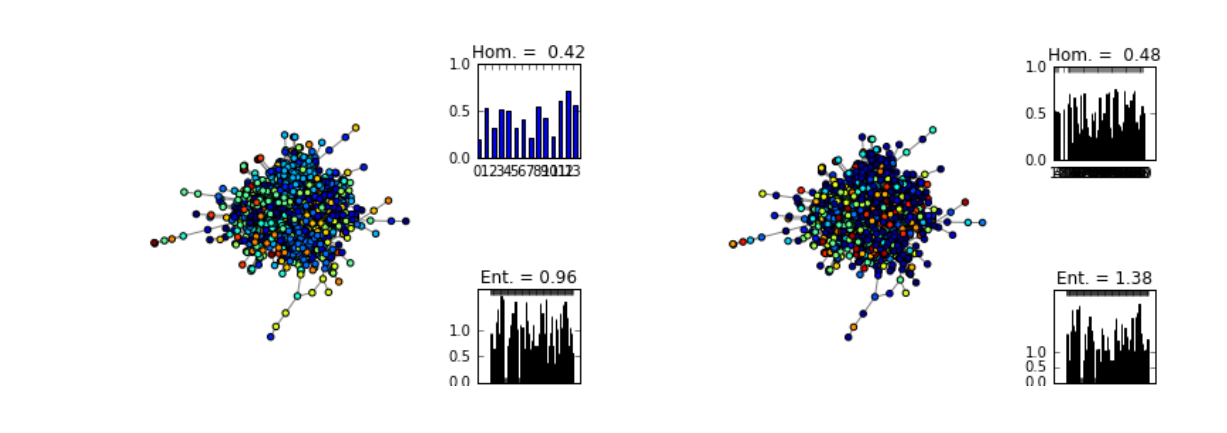}
\caption{Original Louvain partition (left) and finer partition (right) for the painter network.}
\label{paintres}
\end{figure}

We observe in Figure \ref{paintres} that the standard implementation of the Louvain method (producing 14 communities) is a good balance between homogeneity and entropy; however as the number of communities is too small (underdetection) we wish to perform some community detection within the clusters to obtain a finer partition.

\begin{figure}
\centering
\includegraphics[width=0.5\textwidth]{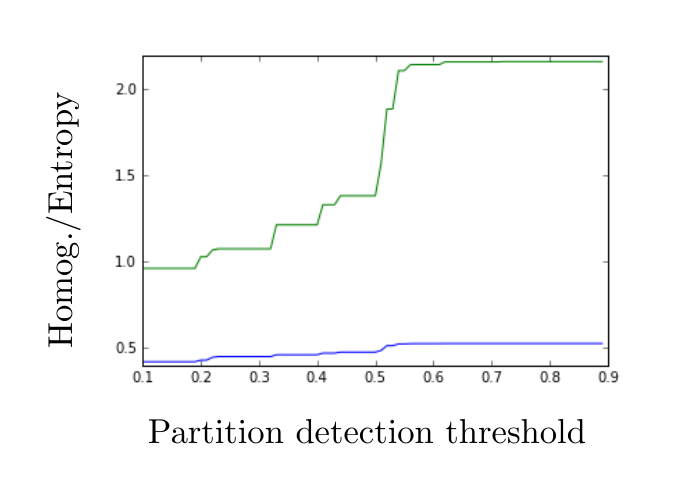}
\caption{Homogeneity and entropy for varying partition detection thresholds.}
\label{threshold}
\end{figure}

As some of the clusters in the partition score highly in homogeneity, we do not need to look into the structure of all; we set a critical homogeneity threshold, below which a cluster with that value of homogeneity will be split into sub-clusters. In Figure \ref{threshold} we see that an appropriate value without compromising entropy is around 0.5; this partition has 72 communities. The finer partition is also shown in Figure \ref{paintres} and we will use both community structures in the following section to identify node influence by considering centrality measures.

\section{Identifying influential nodes}\label{scentrality}

One of our main objectives in this work is to identify influential painters or, conversely, those whose work is influenced from a large number of sources. While this can be answered plainly using a wide range of centrality measures, we propose ways of identifying influential nodes by taking an underlying community structure into account.

We define some preliminary notation. Given a community partition $\mathcal{C} = \left\lbrace c_{\alpha},\ldots, c_{\nu}\right\rbrace$, we denote by $c(i)$ the cluster into which the $i$-th node belongs. Then we can define the kronecker-delta $\delta_{c(i),c(j)}$, which is 1 if nodes $i$ and $j$ are in the same community, and 0 otherwise.

\subsection{Mixing parameter}

\begin{figure}

\includegraphics[width=\textwidth]{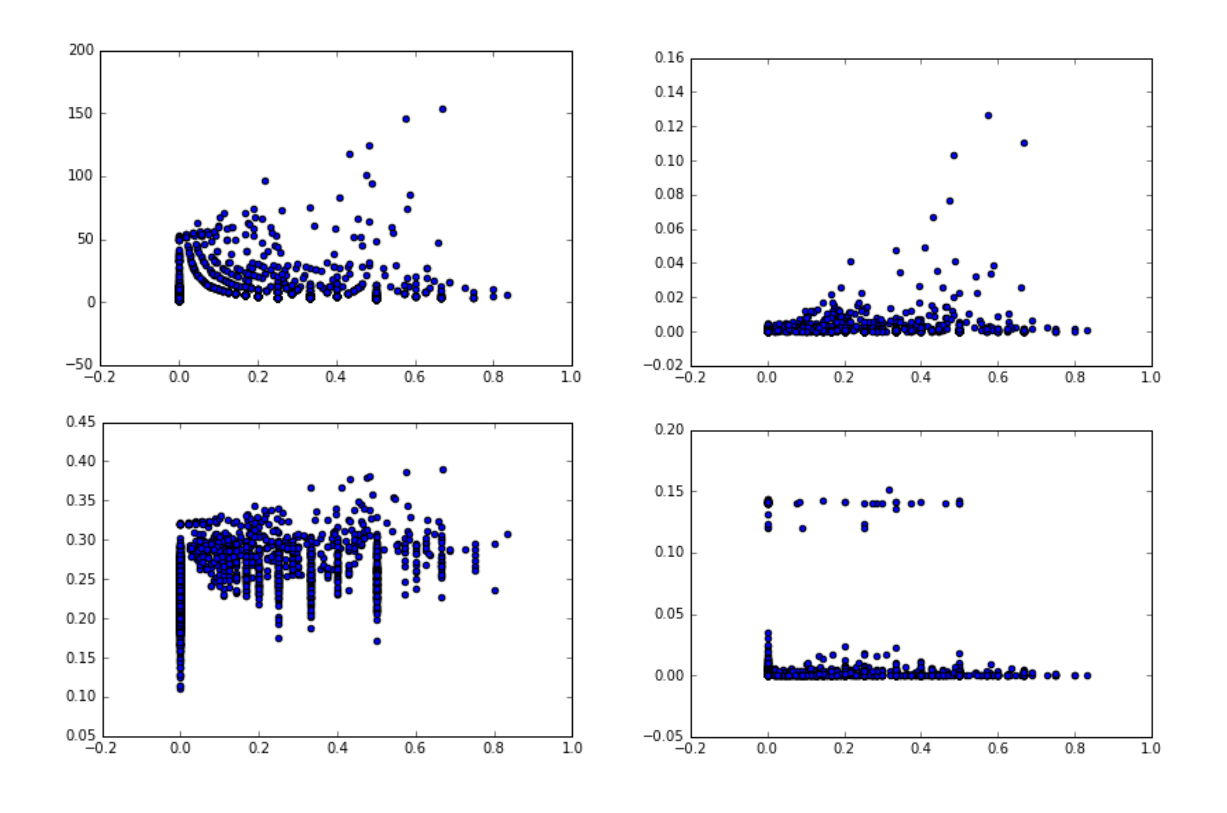}
  \caption{Correlation of $\mu_{\mathcal{C}}(i)$ with centrality measures (clockwise from top left: degree, betweenness, eigenvector and closeness centrality).}
  \label{fig7} 
\end{figure}

Splitting a node's degree given a community structure into links within the community and links outside the community, respectively $k_i^{in}$ and $k_i^{out}$, is commonly occurring in the literature. We propose looking first at the ratio of the outward connections, which is also known in the literature as the \textit{mixing parameter} \citep{Fort16}, given by
\begin{equation}
\mu_{\mathcal{C}}(i) = \frac{k_i^{out}}{k_i}
\end{equation}
Figure \ref{fig7} shows the correlation of this measure with standard centrality measures; the correlation is relatively weak, which illustrates that this measure can indeed have a significant contribution in highlighting nodes which the other measures may not identify.

\subsection{Community-based betweenness centrality}

In order to generalise betweenness centrality, we define the \textit{community-based betweenness centrality} (CBBC), where we wish to only take into account paths that start and finish in different communities.
\begin{equation}
bc_{\mathcal{C}}(i) = \sum_{k,l: \delta_{c(k),c(l)}=0} \frac{\sigma_{kl}(i)}{\sigma_{kl}}
\end{equation}

\begin{figure}
\includegraphics[width=\textwidth]{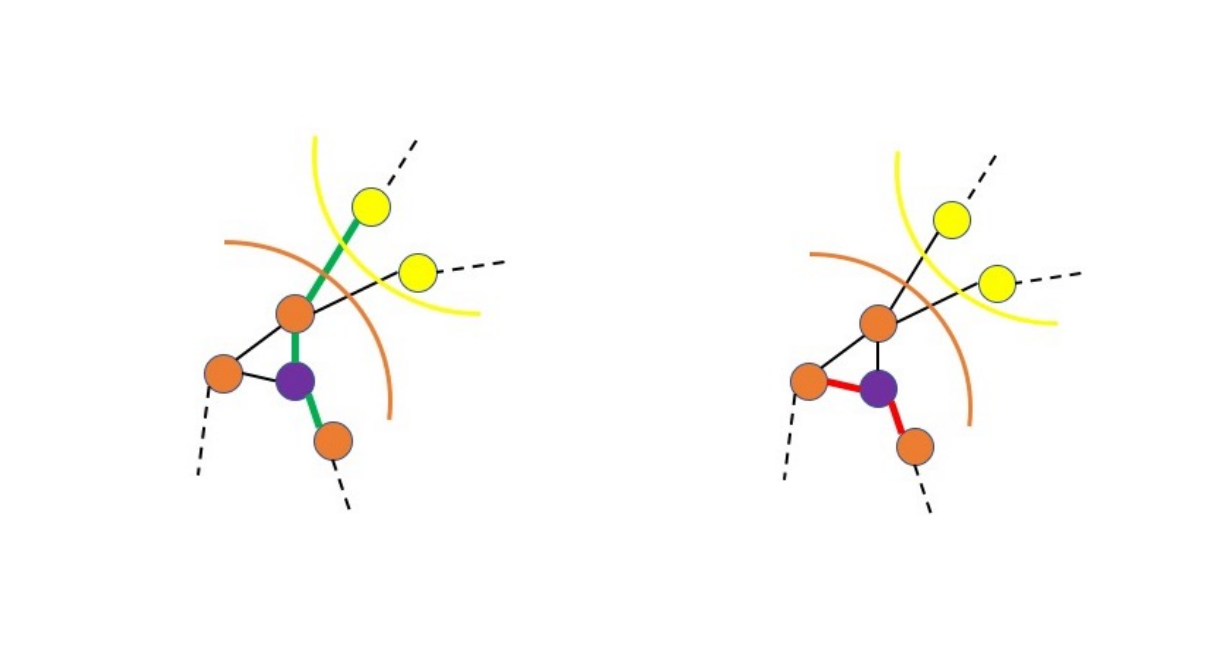}
\caption{Visualisation of the paths taken into account for the CBBC of the purple node. Green path (left) contributes to CBBC, red path (right) does not.}
\label{cbbc}
\end{figure}

A visualisation of this definition can be seen in Figure \ref{cbbc}. The intuition behind using this measure is that an influential painter according to our understanding, is one who promotes the flow of ideas to different communities. As a result we are interested in their position along a shortest path starting and finishing in different clusters.

The correlations between the standard and modified betweenness centrality are quite high for both of our partitions (almost 1). However the ranks of the nodes exhibit smaller correlation values (around 0.88) allowing us to identify certain nodes who score poorly in the standard Betweenness Centrality and better in our Community-Based modification (Figure \ref{cbbc}).

\begin{figure}
\centering
\includegraphics[width=\textwidth]{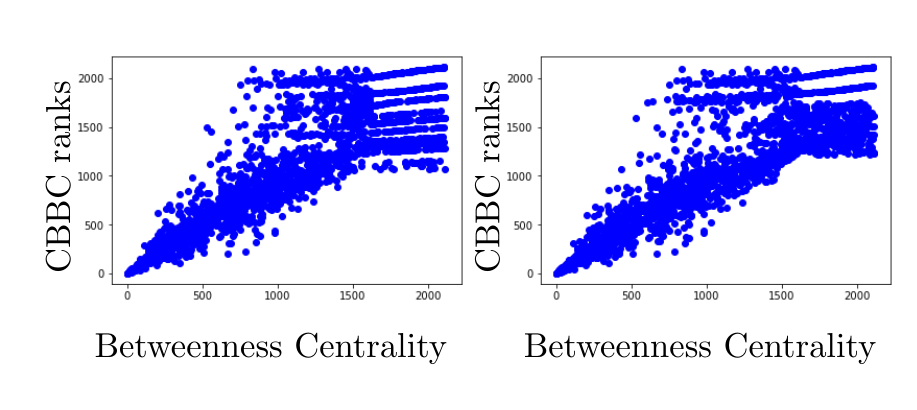}
\caption{Correlations between standard and community-based betweenness centrality, in the original (left) and fine (right) partitions.}
\end{figure}

\subsection{Community-based closeness centrality}
 
Very similar in concept to the betweenness centrality, we define the \textit{community-based closeness centrality} (CBCC) on a node, only considering the shortest distances of nodes in other communities than the node.

\begin{equation}
cc_{\Ccal}(i) = \frac{1}{\sum_{j\notin c(i)}d_{ij}}
\end{equation}

\begin{figure}
\centering
\includegraphics[width=0.35\textwidth]{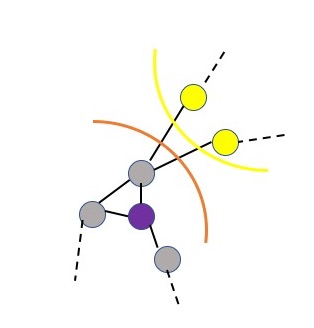}
\caption{Visualisation of the CBCC; only the yellow nodes' distances from the purple node are taken into account to define its score.}
\label{cbcc}
\end{figure}

A visualisation of this generalised centrality measure is on Figure \ref{cbcc}. Correlations are again quite high between the standard and modified centrality measures, though for the finer partition the correlation value is smaller (0.86 against 0.92 for the original partition) which enables us to highlight more painters scoring highly in the modified measures who wouldn't be highlighted in the standard ones.

\begin{figure}
\centering
\includegraphics[width=\textwidth]{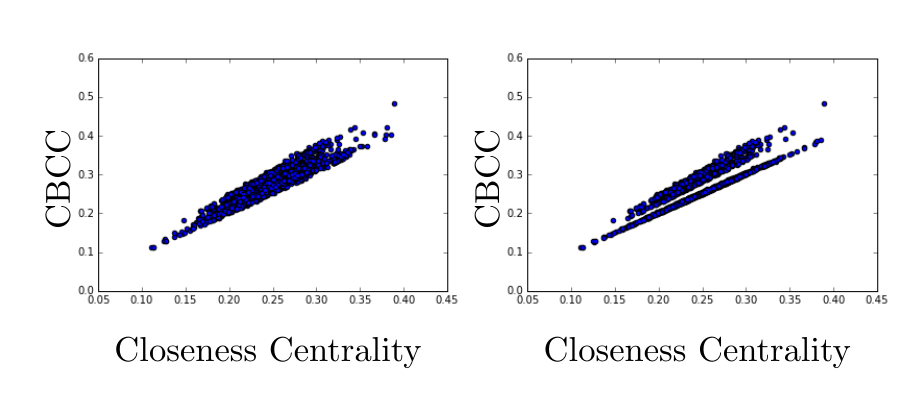}
\caption{Correlations between standard and community-based closeness centrality, in the original (left) and fine (right) partitions.}
\end{figure}

\section{Conclusion}

In this work we have introduced a new context where the theory and methodology of networks based on contextual and biographical links can be applied, by constructing and analysing the network of Western art painters. Our overall aim was to use the properties of the network to highlight painter-nodes who can be classified as influential in art history.

It became clear through our analysis that clusters, as generated by a modularity maximisation algorithm, correspond broadly to artistic movements and areas where painters were active. Motivated by this observation we proposed two measures for assessing a partition into communities, using metadata for artistic movement and location, in order to assess the resolution of a community partition. We have applied this approach to two stylised networks as well as for the analysis of the painter network.

In order to illustrate the nodes with influence and connections beyond their artistic movement and region, we have proposed looking at centrality measures which take community structure into account, and look into the outward links that a node may have. The redefined centrality measures in terms of the community structure are then used to highlight influential nodes who might have been missed as they don't necessary rank too highly in the standard measures. Some examples can be found in the Supplementary Information.

\bibliographystyle{apalike}
\bibliography{Community_Detection_with_Metadata_in_a_Network_of_Biographies_of_Western_Art_Painters.bbl}

\pagebreak

\appendix

\begin{center}
{\Large\textbf{Supplementary Information}} \\[0.5cm]
 {\large {Michael Kitromilidis},
  \href{http://www.imperial.ac.uk/people/t.evans}{Tim S.\ Evans}}
 \\[0.5cm]
\href{http://complexity.org.uk/}{Centre for Complexity Science}, and \href{http://www3.imperial.ac.uk/theoreticalphysics}{Theoretical Physics Group},
Imperial College London, SW7 2AZ, U.K.
\end{center}

\section{Details about the community partition}

The Louvain modularity maximisation method in its standard implementation reveals 14 communities in the painter network (Table \ref{comms} displays the 12 significant ones).

\begin{table}[H]
\centering
\caption{Significant Louvain communities of painters (number of mentions in brackets).}
\begin{tabular}{p{10mm} p{10mm} p{40mm} p{40mm} p{40mm}}
\hline
\textbf{Tag} & \textbf{Size} &  \textbf{Notable Artists} & \textbf{Movements} & \textbf{Locations} \\
\hline
\hline
\#0 & 171 & Turner, Delacroix & Romanticism (75) & French (32) \\ & & & & English (31)  \\ 
\hline
\#1 & 301 & Poussin, Caracci & Baroque (212) & Italian (221) \\ & & &  Rococo (39) \\
\hline
\#2 & 136 & D\"urer, van Mander & North. Renaissance (80) & Flemish (60)\\
\hline
\#3 & 436 & Rubens, van Dyck, Breughel & Baroque (353) & Dutch (200) \\ & & & Mannerism (39) & Flemish (166) \\
\hline
\#4 & 261 & Raphael, Da Vinci, Vasari & High Renaissance (81) & Italy (230) \\ & & & Mannerism (63) \\
\hline
\#5 & 201 & Monet, C\'ezanne, Manet & Impressionism (129) & French (76) \\ & & & Realism (48)  \\
\hline
\#6 & 262 & David, Ingres & Baroque (61) & French (149) \\ & & & Neoclassicism (48)  \\ & & & Rococo (46)  \\
\hline
\#7 & 137 & Titian, Tintoretto & Baroque (65) & Spanish (33) \\ & & &  Mannerism (22)  \\
\hline
\#8 & 50 & - & Baroque (37) & Dutch (35) \\
\hline
\#9 & 130 & Rembrandt, Caravaggio & Baroque (107)
 & Italian (38) \\ & & & & Dutch (27)\\
\hline
\#10 & 52 & - & Realism (18) & Hungarian (16) \\ & & & Romanticism (16)  \\
\hline
\#11 & 30 & - & Realism (20) & Italian (26) \\
\hline
\hline
\end{tabular}
\label{comms}
\end{table}

We note that some communities (e.g. 1, 3, 5 and 9) are movement-based, whereas others (e.g. 4, 6) are mainly location-based.

\section{Metadata for painters}

We collect the main artistic movements as identified in the WGA and also what we observe from our analysis in Section 3. The tags associated with each node in terms of their location and artistic movement are as follows:

\paragraph{Movement} Medieval, Early Renaissance, Northern Renaissance, High Renaissance, Mannerism, Baroque, Rococo, Neoclassicism, Romanticism, Realism, Impressionism

\paragraph{Location} American, Austrian, Belgian, Bohemian, Catalan, Danish, Dutch, English, Finnish, Flemish, French, German, Greek, Hungarian, Irish, Italian, Netherlandish, Norwegian, Polish, Portuguese, Russian, Scottish, Spanish, Swedish, Swiss

\section{Examples highlighted by the new measures}

Some examples of painters highlighted by the modified centrality measures are as follows:

\subsection{Mixing parameter}

The mixing coefficient is likely to highlight nodes which score poorly on other centrality measures due to its relatively low correlations. This measure highlights among others the artist \textbf{Jean-Baptiste-Sim\'eon Chardin}.

Chardin is placed in community \#5, which is mainly a cluster with Impressionists. He was actually living in 18th century France but was very influential for the much later impressionist artists, who as a result are frequently linked to him. As he naturally has connections to his contemporary artists, mainly placed in community \#6, this gives him a high mixing coefficient ($\mu = 0.69$), even though his degree ($k=16$) is comparatively low.

\begin{figure}[H]
\centering
\begin{subfigure}[l]{0.35\textwidth}
\includegraphics[width=1.05\textwidth,height=40mm]{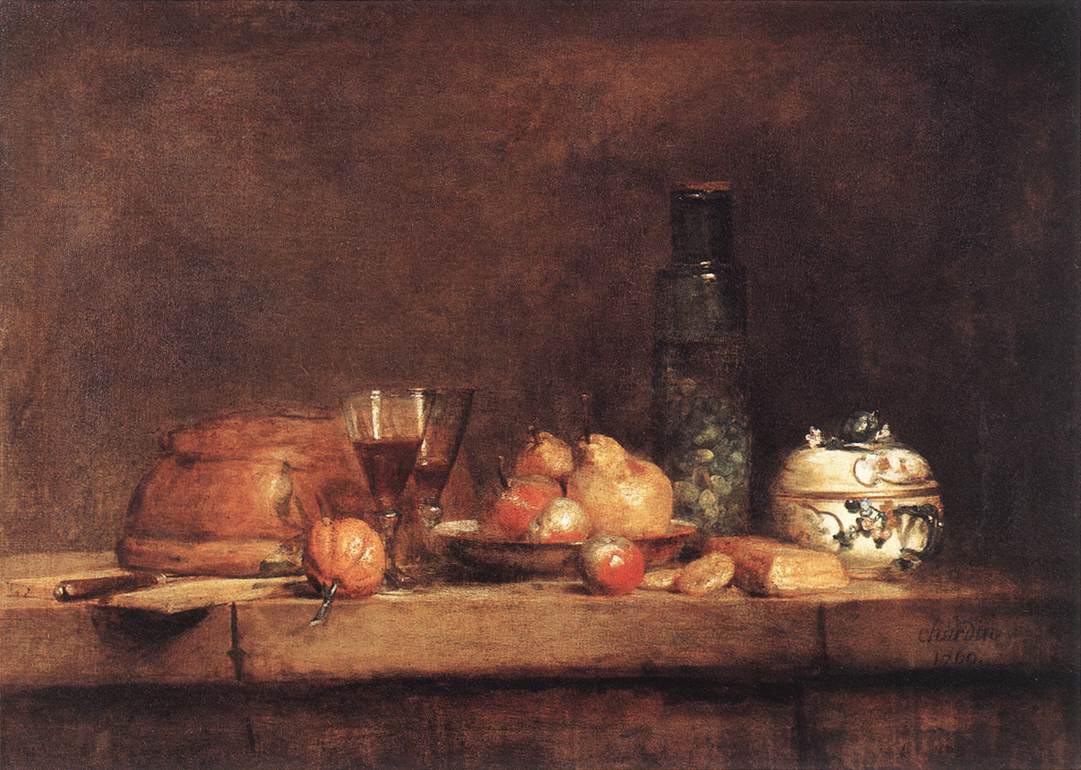}
\subcaption{Chardin (1728)}
\end{subfigure}
\hspace{10mm}
\begin{subfigure}[r]{0.35\textwidth}
\includegraphics[width=1.05\textwidth,height=40mm]{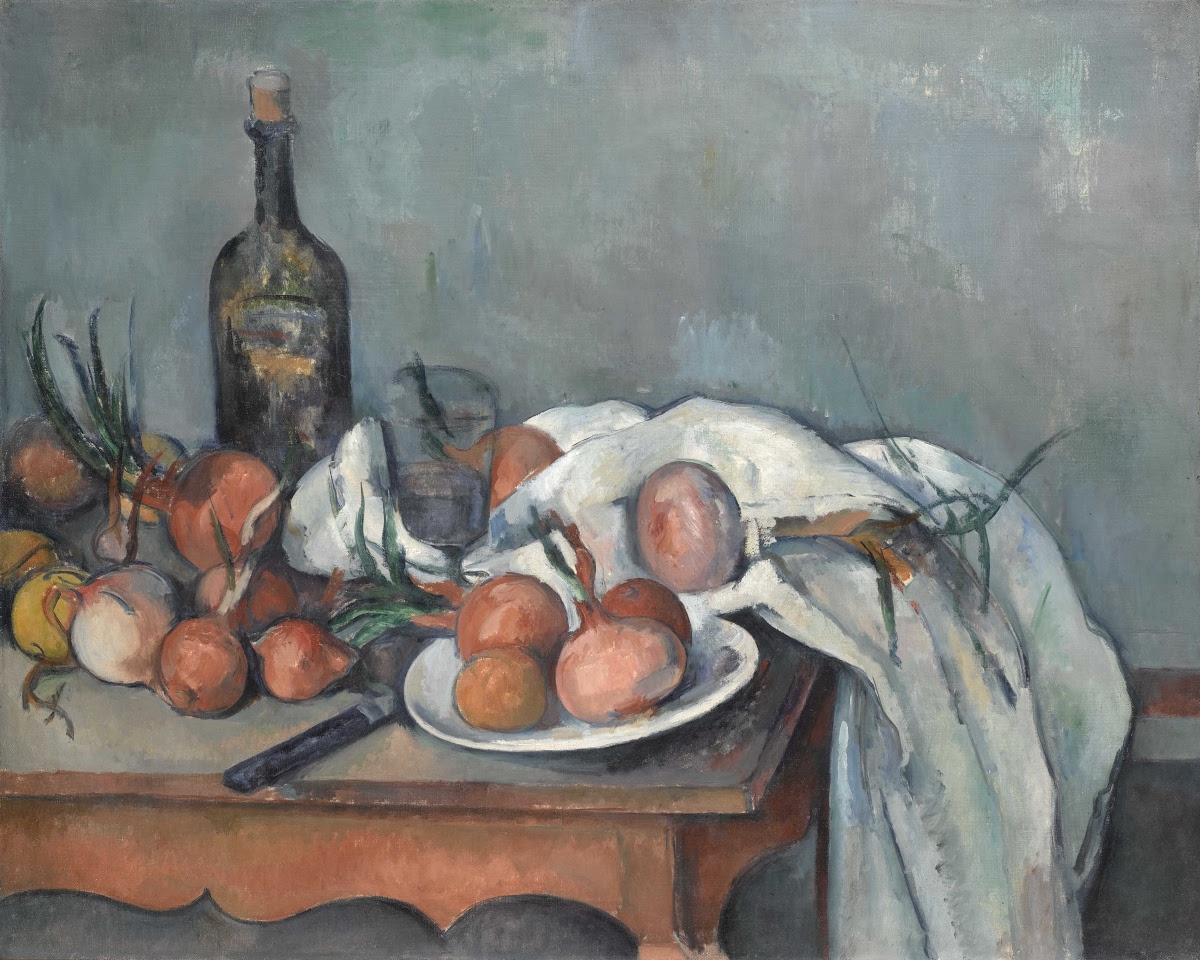}
\subcaption{C\'ezanne (1898)}
\end{subfigure}
\caption{Example of how Chardin's still life works were highly influential for impressionists, such as C\'ezanne.}
\end{figure}

\subsection{Community-based betweenness centrality}

An example of an artist whose betweenness centrality rank improves when communities are taken into account is the German-Jewish painter \textbf{Max Liebermann}. In terms of standard betweenness centrality he is ranked slightly lower than 300th, whereas in the modified measures his position improves by around 100 slots, both in the coarse and in the fine partition (with a slightly better increase in the fine partition indeed).

Liebermann was a key figure in bringing impressionism from France to Germany, as he had a large personal collection of French impressionist paintings, and together with contemporary artists helped shape the movement in his country. As a result, we expect that he will appear along paths linking French impressionists to German artists of a slightly later period, thus boosting his community-dependent score.

\begin{figure}[H]
\centering
\begin{subfigure}[l]{0.35\textwidth}
\includegraphics[width=1.05\textwidth,height=40mm]{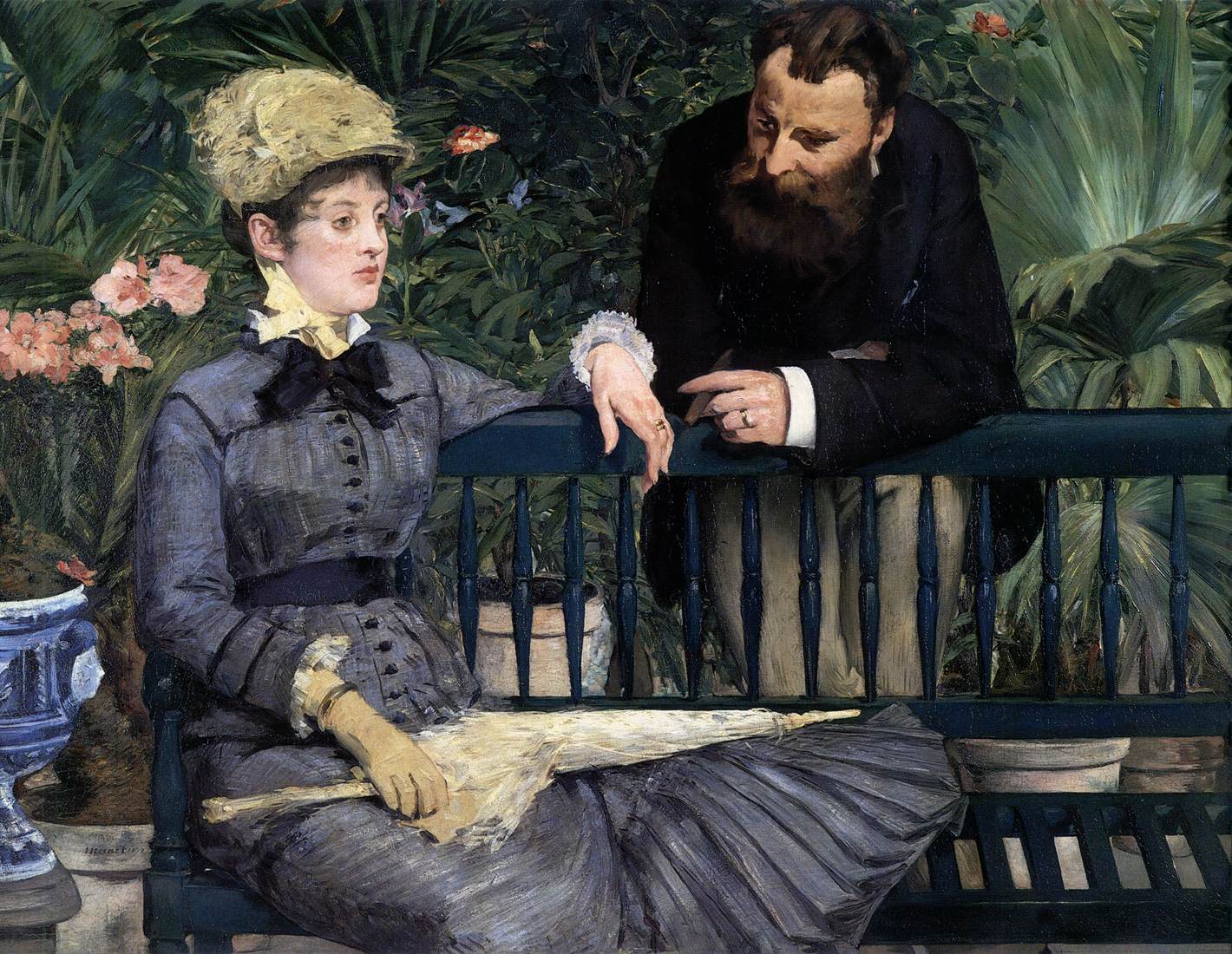}
\subcaption{Manet (1879)}
\end{subfigure}
\hspace{10mm}
\begin{subfigure}[r]{0.35\textwidth}
\includegraphics[width=1.05\textwidth,height=40mm,width=35mm]{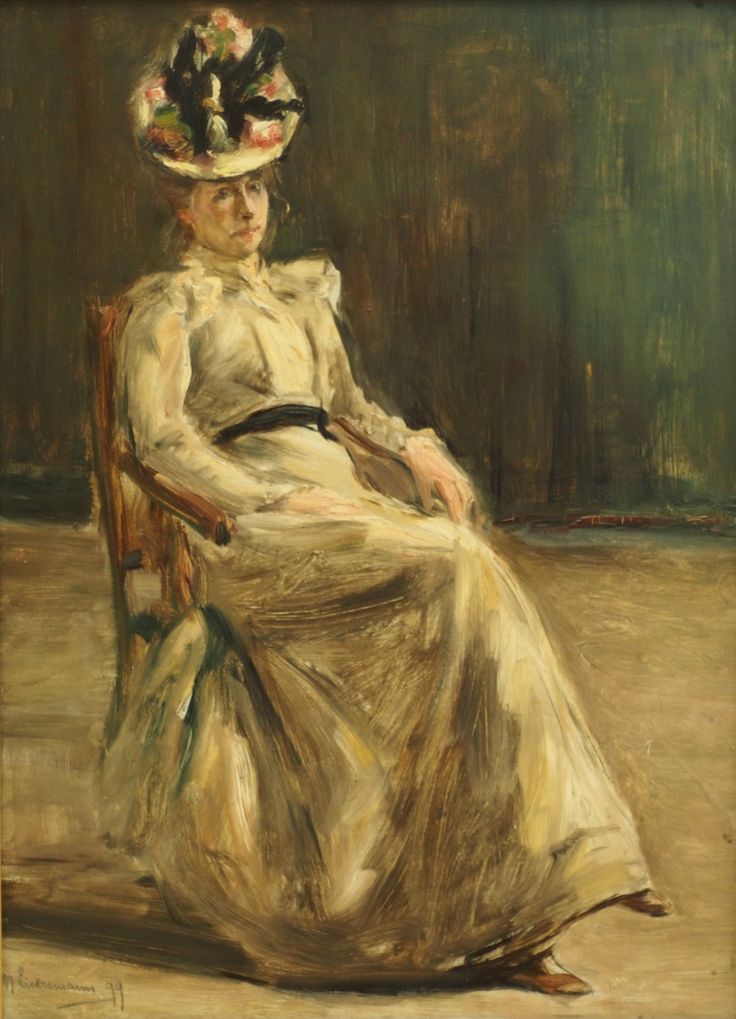}
\subcaption{Liebermann (1894)}
\end{subfigure}
\caption{Liebermann was influenced by French impressionists such as Manet, whom he had encountered during his stay in Paris; here we see him adopting the more relax posture for his \textit{Portrait of a Seated Lady} from Manet's \textit{Winter Garden}.}
\end{figure}

\subsection{Community-based closeness centrality}

Although strongly correlated with the standard closeness centrality, CBCC can highlight artists with connections to multiple communities outside their own. One such example is \textbf{Jacob van Ruisdael}, a Dutch Baroque painter. He is appropriately placed in community \#3, and even though he ranks 96th in closeness centrality overall, he is 9th in the modified centrality applied in the finer partition.

As a landscape painter, his style influenced later artists in multiple genres, including English romanticists as well as impressionists. This means that, despite the fact that the majority of his links are with Dutch Baroque artists in the same cluster as him, his community-based closeness is high as he has links to artists in several other communities as well.

\begin{figure}[H]
\centering
\begin{subfigure}[l]{0.35\textwidth}
\includegraphics[width=1.05\textwidth,height=40mm]{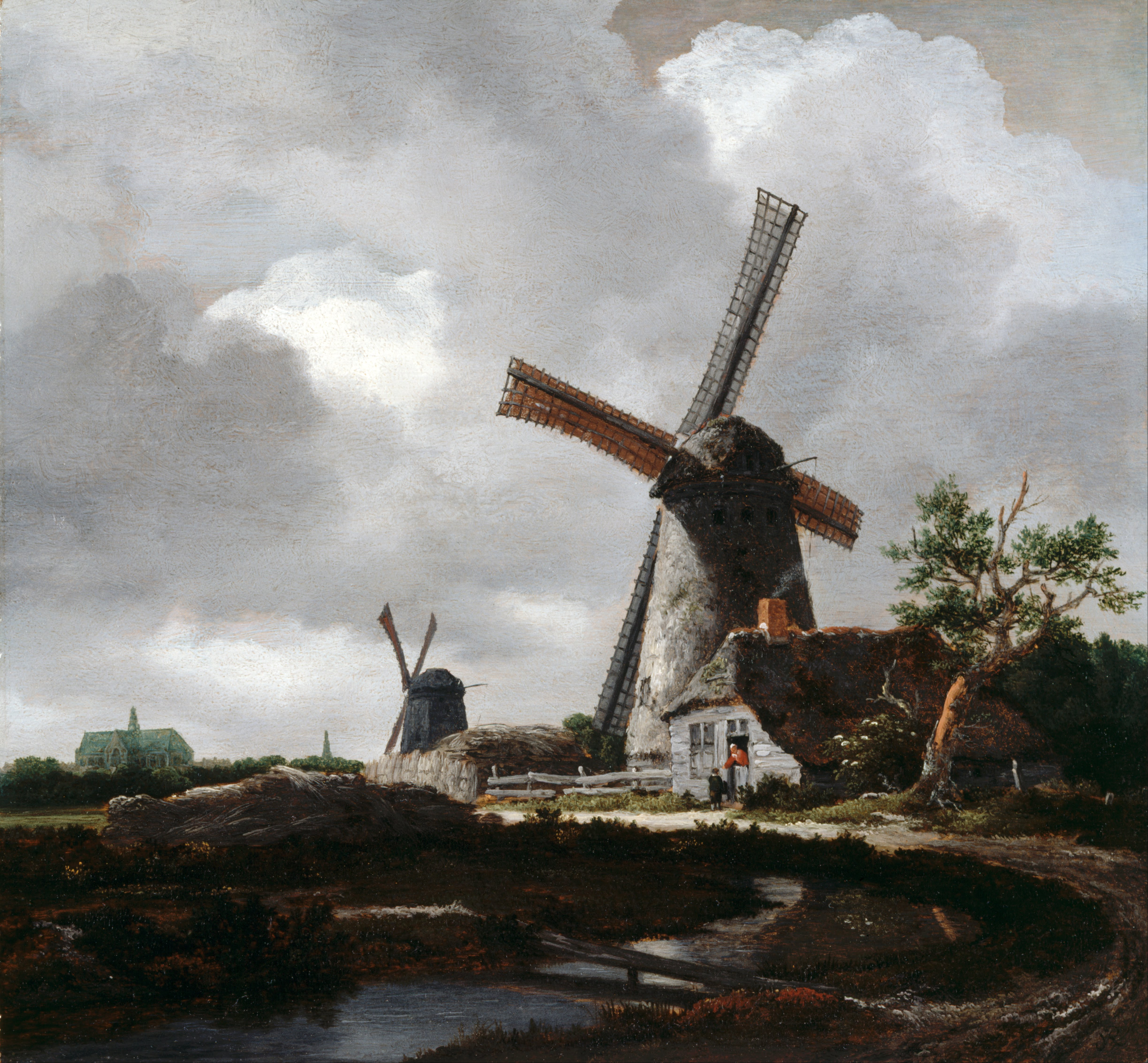}
\subcaption{Ruisdael (1640)}
\end{subfigure}
\hspace{10mm}
\begin{subfigure}[r]{0.35\textwidth}
\includegraphics[width=1.05\textwidth,height=40mm]{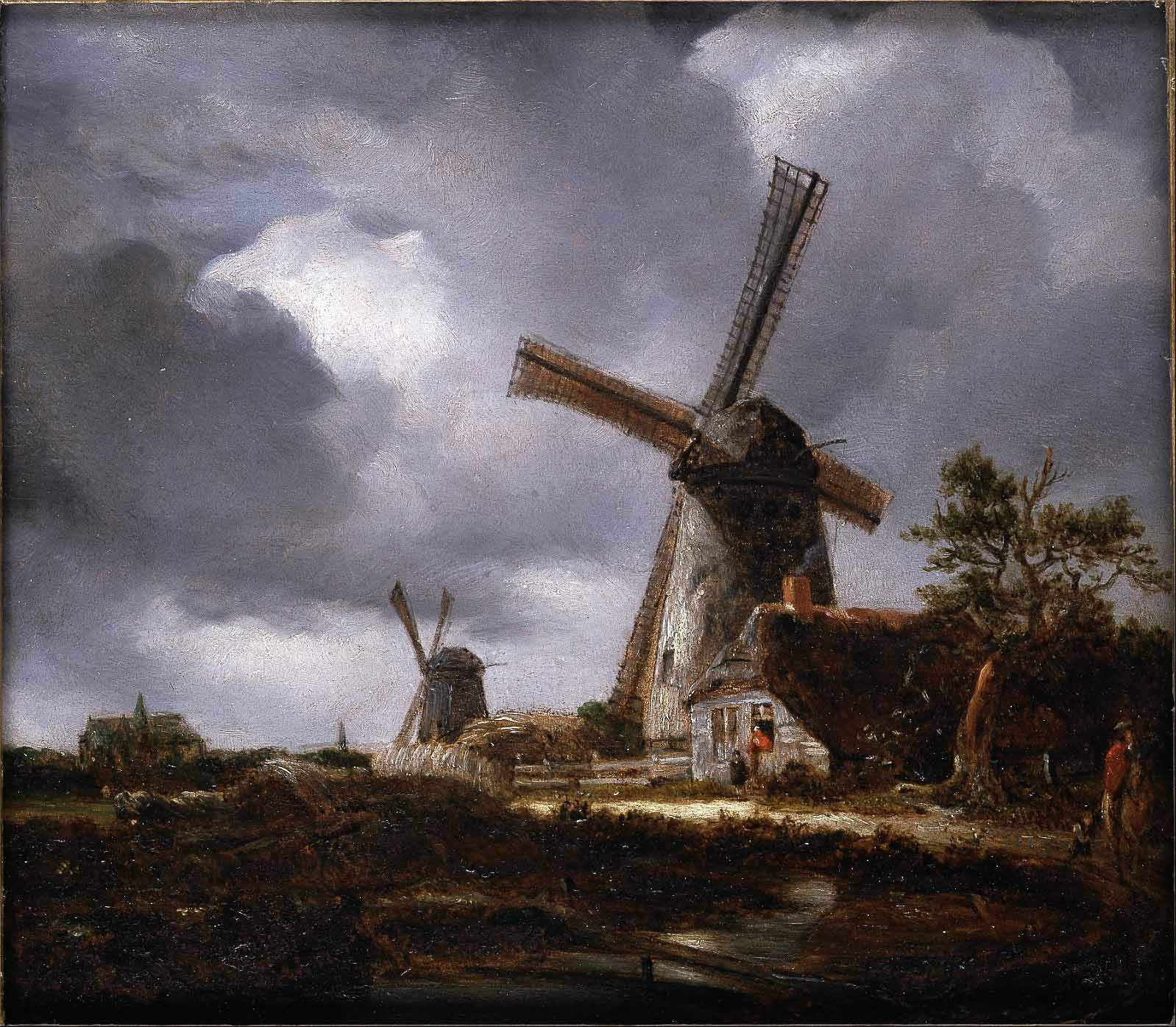}
\subcaption{Constable (1830)}
\end{subfigure}
\caption{Ruisdael's \textit{Landscape with windmills} was studied by many artists for his landscape painting techniques, including in this copy by John Constable.}
\end{figure}

\begin{figure}[H]
\centering
\begin{subfigure}[l]{0.35\textwidth}
\includegraphics[width=1.05\textwidth,height=40mm]{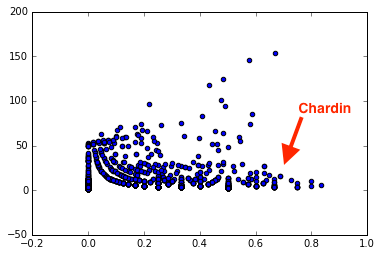}
\end{subfigure}
\hspace{10mm}
\begin{subfigure}[r]{0.35\textwidth}
\includegraphics[width=1.05\textwidth,height=40mm]{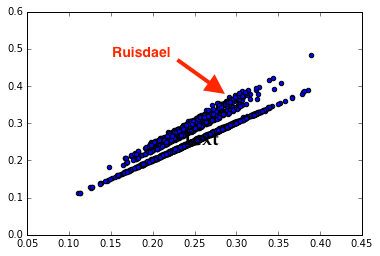}
\end{subfigure}
\caption{Chardin's and Ruisdael's positions in terms of the degree-mixing coefficient (left) and closeness-CBCC (right) correlations.}
\end{figure}

\section{References}
References for the Supplementary Material only:
\begin{itemize}
\item[1] Liebermann, Max. Max Liebermann and International Modernism: An Artist's Career from Empire to Third Reich. Vol. 14. Berghahn Books, 2011.
\item[2] McCoubrey, John W. "The Revival of Chardin in French Still-Life Painting, 1850–1870." The Art Bulletin 46.1 (1964): 39-53.
\item[3] Slive, Seymour. Jacob van Ruisdael: windmills and water mills. Getty Publications, 2011.
\end{itemize}

\end{document}